\documentclass{elsart}

\usepackage{epsfig}
\usepackage{amsmath}
\usepackage{amssymb}
\usepackage{dcolumn}

\begin{document}

\begin{frontmatter}



\title{Limits to the muon flux from neutralino annihilations in the Sun with the AMANDA detector}

\author[DESY]{M.~Ackermann},
\author[Mainz]{J.~Ahrens},
\author[Bartol]{X.~Bai},
\author[Dortmund]{M.~Bartelt},
\author[UCI]{S.~W.~Barwick},
\author[UCB]{R.~C. Bay},
\author[Mainz]{T.~Becka},
\author[Wuppertal]{K.-H.~Becker},
\author[Dortmund]{J.-K.~Becker},
\author[ULB]{P. Berghaus},
% ...
\collab{(The AMANDA collaboration)}

\address[DESY]{DESY, Platanenallee 6, D 15738 Zeuthen, Germany}
\address[Mainz]{Institute of Physics, University of Mainz, D-55099 Mainz, Germany}
\address[Bartol]{Bartol Research Institute, University of Delaware, Newark, DE 19716, USA}
\address[Dortmund]{Institute of Physics, University of Dortmund, D-44221 Dortmund, Germany }
\address[UCI]{Department of Physics and Astronomy, University of California, Irvine, CA 92697, USA}
\address[UCB]{Department of Physics, University of California, Berkeley, CA 94720, USA}
\address[Wuppertal]{Department of Physics, University of Wuppertal, D-42119 Wuppertal, Germany}
\address[ULB]{Universit\'e Libre de Bruxelles, Science Faculty CP230, B-1050 Brussels,~Belgium}
\address[Uppsala]{Division of High Energy Physics, Uppsala University. S-75121 Uppsala, Sweden}
\address[Stockholm]{Department of Physics, Stockholm University, S-10691 Stockholm, Sweden}
\address[Mons]{Faculty of Sciences, University of Mons-Hainaut, B-7000 Mons, Belgium}
\address[UPenn]{Department of Physics and Astronomy, Pennsylvania State University, PA 16802, USA}
\address[UWM]{Department of Physics, University of Wisconsin-Madison, WI 53706, USA}
\address[VUB]{Vrije Universiteit Brussel, Dienst ELEM, B-1050 Brussels,~Belgium}
\address[Maryland]{Department of Physics, University of Maryland, College Park, MD 20742, USA}
\address[LBNL]{Lawrence Berkeley National Laboratory, Berkeley, CA 94720, USA}
\address[IC]{Blackett Laboratory, Imperial College, London SW7 2BW, UK}
\address[Kalmar]{Department of Chemistry and Biomedical Sciences, University of Kalmar, SE-39182 Kalmar, Sweden}
\address[UWRF]{Department of Physics, University of Wisconsin-River Falls, WI 54022, USA}

\begin{abstract}
  A search for an excess of muon-neutrinos from 
neutralino annihilations in the Sun has been performed 
with the AMANDA-II neutrino detector using data collected 
in 143.7 days of live-time in 2001. No excess over the expected atmospheric
neutrino background has been observed. An upper limit at 90\% confidence 
level has been obtained on the annihilation 
rate of captured neutralinos in the Sun, as well as the
corresponding muon flux limit at the Earth, both as functions of the
neutralino mass in the range 100~GeV-5000~GeV.
\end{abstract}


\end{frontmatter}

\section{\label{sec:Intro} Introduction}

The Minimal Supersymmetric extension to the Standard Model of particle
physics (MSSM)~\cite{Martin:97a} provides a promising dark matter candidate in the
lightest neutralino, $\widetilde{\chi}^{\tiny 0}_{\tiny 1}$, a linear combination of the 
 supersymmetric partners of the electroweak neutral gauge- and Higgs
 bosons. Assuming R-parity conservation, the neutralino is stable.  
From accelerator searches and relic density constraints from WMAP data,  
a lower limit on the mass of the MSSM neutralino can be
derived~\cite{Genieveve:04a}. Typical lower limits for $m_{\widetilde{\chi}^{\tiny 0}_{\tiny 1}}$ 
from such studies are about 20 GeV, depending on the values chosen for 
tan$\beta$. In models where the pseudo-scalar Higgs boson $A$ is assumed
to be light, $m_A<$ 200 GeV, the  neutralino mass can be as low as 6 GeV. 
Theoretical arguments based on the requirement of unitarity set an upper limit 
on  $m_{\widetilde{\chi}^{\tiny 0}_{\tiny 1}}$ of 340~TeV~\cite{Griest:90a}. Within these limits, the allowed 
parameter space of minimal super-symmetry can be exploited to build realistic models which provide 
relic neutralino densities of cosmological interest to address the
dark matter problem.\par
 Relic neutralinos in the galactic halo can become gravitationally bound in orbits 
in the solar system by losing energy through elastic scattering with 
matter. They may finally be trapped and accumulate inside celestial bodies like 
the Sun~\cite{Press:85a,Freese:86a,Gould:88a},  annihilate, and produce 
a neutrino flux  from the decays of the annihilation
products~\cite{Jungman:96a}. \par
 We have previously published a search for neutralino dark matter accumulated in the 
Earth using the 10-string detector AMANDA-B10 and data from 1997~\cite{AMANDA:02a}. 
In this letter we present the first search for neutralino dark matter accumulated 
in the Sun with the extended AMANDA-II detector. Due to the position of the Sun at the South Pole, 
reaching no more than 23.5$^\circ$ below the horizon, the separation of a potential neutralino-induced neutrino flux 
from the Sun from the atmospheric muon flux is a challenge that we could only address because of the increased sensitivity of the AMANDA-II detector to horizontal tracks.

\section{\label{sec:detector}The AMANDA-II detector}

 AMANDA-II  consists of an array of 677 optical modules deployed on  
19 vertical strings at depths between 1500~m and 2000~m in the South Pole 
ice cap. The strings are arranged in three approximately concentric 
circles of 60~m, 120~m and 200~m diameter respectively. 
An optical module consists of a photomultiplier tube housed in a glass pressure vessel. 
 Muons from charged-current neutrino interactions near the array are 
detected by the Cherenkov light they produce when traversing the
ice. During 2001 the detector was triggered when at least 24 modules were hit within a time 
window of 2.5 $\mu$s. The trigger rate was 70~Hz. 
 The relative timing of the  photons reaching the optical modules allows the
reconstruction of the muon track. A more detailed description of the reconstruction techniques used in AMANDA 
is given in Ref.~\cite{AMANDA:04a}. 

\section{\label{sec:simulations} Signal and background simulations}
 The simulation of the neutralino-induced neutrino signal was performed using the {\texttt{DARKSUSY}} 
program~\cite{DARKSUSY} for a sample of neutralino masses (100, 250, 500, 1000, 3000 and 5000 GeV).
Neutralinos can annihilate pair-wise to, e.g., $\ell^{+} \ell^{-}$,
$\mbox{q}\bar{\mbox{q}}$, W$^{+}$W$^{-}$, Z$^0$Z$^0$, H$^0_{1,2}$H$^0_{3}$,
Z$^0$H$^0_{1,2}$ and W$^{\pm}$H$^{\mp}$, and neutrinos are produced in the decays of the annihilation
products. Neutrinos produced in quark jets (from e.g.\
$\mbox{b}\bar{\mbox{b}}$) or from Higgs bosons typically have lower energy than those produced from
decays of $\tau$ leptons or gauge bosons. Two annihilation channels
were considered in this paper for each mass tested, $\widetilde{\chi}^{\tiny 0}_{\tiny 1}\widetilde{\chi}^{\tiny 0}_{\tiny 1}\rightarrow \mbox{b}\bar{\mbox{b}}$
and $\widetilde{\chi}^{\tiny 0}_{\tiny 1}\widetilde{\chi}^{\tiny 0}_{\tiny 1}\rightarrow$ W$^{+}$W$^{-}$,
which are referred to as {\it soft} and {\it hard}, respectively, in the rest of
the paper. This choice covers the range of neutrino energies which would be detectable with AMANDA for typical
MSSM models. The simulated angular range was restricted to zenith angles between 90$^\circ$ (horizontal)
and 113$^\circ$, with the generated number 
of events as a function of angle weighted by the time the Sun spends at each declination.
The neutrino-nucleon interactions in the ice around the detector producing the detectable muon 
flux were simulated with {\texttt{PYTHIA}}~\cite{PYTHIA} using the {\texttt{CTEQ3}}~\cite{CTEQ3} 
parametrization of the nucleon structure functions.\par

The background for this search arises from up-going atmospheric neutrinos and mis-reconstructed downward-going 
atmospheric muons. We have simulated the atmospheric neutrino flux according to~\cite{Lipari:93a}. The simulation includes neutrino
propagation through the Earth, taking into account the Earth density
profile, neutrino absorption and neutral current scattering. 
 A sample of 1.9$\times 10^{4}$ atmospheric neutrinos with energies between
10~GeV and 10~PeV and zenith angles between 70$^\circ$ 
and 180$^\circ$ (vertically up-going) has been simulated.\par
The simulation of atmospheric muons was based on the {\texttt{CORSIKA}} air shower generator~\cite{CORSIKA} 
using the South Pole atmosphere parameters and protons as primaries. We have simulated 1.6~$\times$~$10^{8}$  
interactions, distributed isotropically with zenith angles between 0$^\circ$ and 85$^\circ$,
and with proton energies, $E_p$,  between 600~GeV and 10$^{11}$~GeV, assuming a differential energy 
distribution $\propto E_p^{-2.7}$~\cite{PDG:00}. The
sample corresponds to about 32.5 days of equivalent detector live-time. \par

Muons were propagated from the production point to the  detector 
taking into account energy losses by bremsstrahlung, pair production, photo-nuclear 
interactions and $\delta$-ray production as implemented in the code {\texttt{MMC}}~\cite{MMC}.

\begin{figure}[t]
\centering\epsfig{file=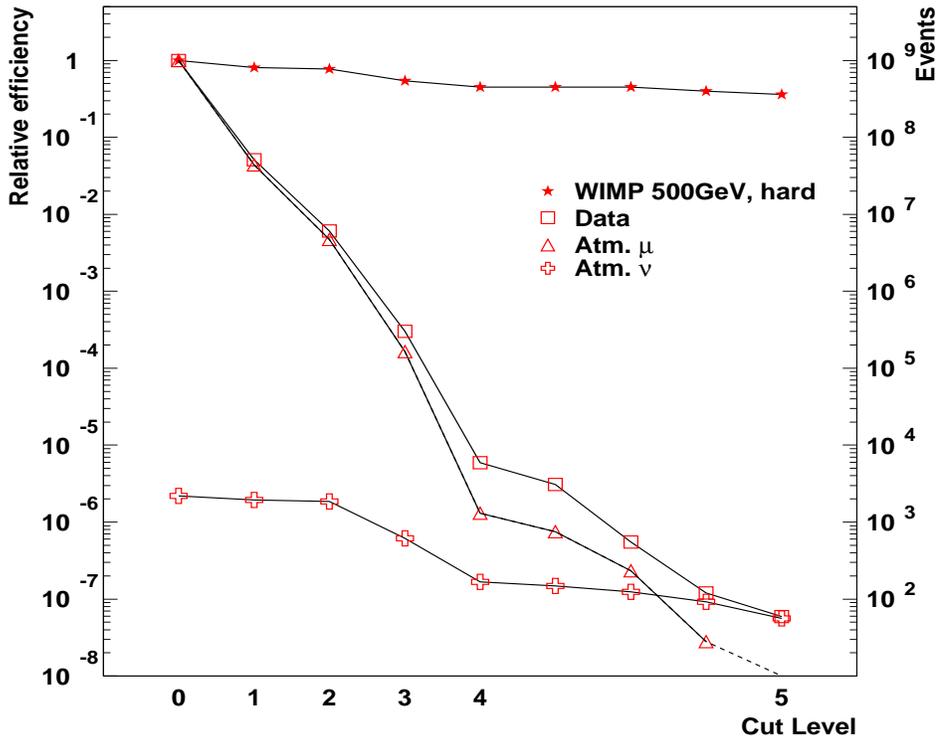,height=0.8\linewidth,width=0.9\linewidth}
\caption{Relative cut efficiencies for one of the neutralino masses and annihilation channels 
considered in this paper, 500 GeV hard. The efficiency is set to 1 at trigger level, except for the
atmospheric neutrino curve, that has been scaled relative to the number of 
events expected in the live-time of the analysis. 
The axis scale on the right gives the absolute normalization 
for the data and the atmospheric neutrino curves.}
\label{fig:relative_effs_L5}
\end{figure}

\section{\label{sec:analysis}Data analysis}

 The data set used comprises 8.7$\times$10$^8$ events collected in 143.7 
 days of effective live-time  between the end of March and the end of September 2001.
The data and Monte Carlo events have been reconstructed with a fast first-guess track
 finding algorithm which is used as a seed for an iterative
 maximum-likelihood reconstruction. 
The reconstructed events were processed through a series of filters in order to
reduce the atmospheric muon background and retain as much of
a potential signal as possible. The variables used in the analysis are the same for the different neutralino 
masses, but the cuts have been optimized independently for each mass
in order to exploit the differences in the resulting muon energy 
spectra. The event selection was optimized in a 'blind' way, by explicitly excluding
 the time tag of the events (and thus the direction of the Sun) in the development of the
 selection criteria (see~\cite{Yulia} for details of this analysis). \par
 Firstly, events were selected according  to the quality of the track reconstruction and their zenith angle. 
The quality criteria were based on  'direct hits', {\em i.e.}
hits with time residuals, $t_{\it res}$, in the interval 
$[-15{\rm ns}, 25{\rm ns}]$ relative to the expected arrival time of an unscattered
photon from the fitted track. Cuts were made on the number of direct hits, the
maximum distance along the track direction between two such hits and the number of
strings with direct hits. 
This selection (cut levels 1 to 3 in figure~\ref{fig:relative_effs_L5}) resulted in a rejection of
99.97\% of the data (mostly downgoing muons), while the simulated background from
atmospheric neutrinos was reduced to 30\%. The simulated signal was reduced between
46\% and 54\% (depending on the neutralino mass). \par

\begin{table}[ht]
\caption{\label{tab:events_at_L5} For each neutralino mass and annihilation channel the 
table shows: The number of data events, N$_{\mbox{\tiny data}}$,  the number of expected
 atmospheric neutrinos in the declination band of
 the Sun, N$^{\mbox{\tiny MC}}_{\mbox{\tiny atm}\,\tiny \nu}$, the  search bin, $\Psi$,  
the final number of data events and the estimated background in the search
 bin,  N$^{\Psi}_{\mbox{\tiny data}}$ and N$^{\Psi}_{\mbox{\tiny bck}}$ respectively and the muon
 effective volume, V$_{\mbox{\tiny eff}}$.  The last three 
columns show 
the 90\% CL upper limit on the expected signal, $\mu_{90}$, and the corresponding 90\% CL limits on the 
annihilation rate at the center of the Sun and on the muon flux at the
  Earth, $\Gamma_{\mbox{\tiny A}}$ 
    and $\Phi_{\mu}$. The limits include systematic uncertainties. 'Soft' and 'hard'
 refer to annihilation into  $b\bar{b}$ and $W^+W^-$ respectively.}
\begin{minipage}{0.8\linewidth}
\begin{tabular}{ccccccccccc}  
  m$_{\widetilde{\chi}^{\tiny 0}_{\tiny 1}}$ & channel & N$_{\mbox{\tiny data}}$ & N$^{\mbox{\tiny MC}}_{\mbox{\tiny atm}\,\tiny \nu}$ &
  $\Psi$&
  N$^{\Psi}_{\mbox{\tiny data}}$ & N$^{\Psi}_{\mbox{\tiny bck}}$ & V$_{\mbox{\tiny eff}}$ &
 $\mu_{90}$ & $\Gamma_{\mbox{\tiny A}}$
    & $\Phi_{\mu}$   \\[-0.3cm]
{\scriptsize{(GeV)}}& & & & {\scriptsize{(deg)}}& & & {\scriptsize (m$^3$)} & &{\scriptsize (s$^{-1}$) }& {\scriptsize (km$^{-2}$ y$^{-1}$)}\\ \hline\hline
  100    & hard  &39  & 42.7 & 11& 1 & 2.5 & 4.2$\times$10$^{4}$& 2.3 &  8.6$\times$10$^{23}$ & 2.5$\times$10$^4$\\
         & soft   & 41 & 45.9 & 26 & 6 & 5.6  & 1.7$\times$10$^{3}$& 9.7 & 8.0$\times$10$^{26}$ & 2.8$\times$10$^6$\\\hline
  250    & hard  & 49  & 46.6 & 8 & 2 & 2.2 & 7.0$\times$10$^{5}$& 4.1 & 5.1$\times$10$^{22}$ & 6.2$\times$10$^3$\\
         & soft   & 53 & 49.4 & 10 & 4 & 2.7 & 5.3$\times$10$^{4}$& 8.9 & 5.5$\times$10$^{24}$ & 7.2$\times$10$^3$\\ \hline
  500    & hard  & 51   & 48.5 & 7 & 1 & 2.0 & 1.7$\times$10$^{6}$& 2.5 & 1.1$\times$10$^{22}$ & 2.4$\times$10$^3$\\
         & soft   & 51 & 50.1 & 8 & 1 & 2.3 & 1.9$\times$10$^{5}$& 2.5 & 2.8$\times$10$^{22}$ & 8.0$\times$10$^2$\\ \hline
  1000    & hard  & 50  & 49.3 & 6 & 1 & 1.7 &2.8$\times$10$^{6}$& 2.9 & 8.5$\times$10$^{21}$ & 2.2$\times$10$^3$\\
         & soft   &51 & 52.2 & 8 &  1& 2.3 & 3.5$\times$10$^{5}$& 2.5 &  1.1$\times$10$^{23}$ & 5.4$\times$10$^3$\\ \hline
  3000    & hard  & 48  & 49.3 & 5.5 & 1 & 1.5 & 2.8$\times$10$^{6}$& 3.1 & 1.4$\times$10$^{22}$ & 2.3$\times$10$^3$\\
         & soft  &49 & 49.7 & 7 & 1 & 1.9  & 4.8$\times$10$^{5}$& 2.7 & 5.5$\times$10$^{22}$ & 4.7$\times$10$^3$\\ \hline
  5000    & hard  & 51   & 47.8 & 5.5 & 1 & 1.6 & 2.6$\times$10$^{6}$& 2.9& 1.9$\times$10$^{22}$ & 2.3$\times$10$^3$\\
         & soft  & 51 & 49.4 & 7 & 1 &  2.0& 5.7$\times$10$^{5}$& 2.7 & 4.3$\times$10$^{22}$ & 4.2$\times$10$^3$\\ \hline\hline
\\
\end{tabular}
\end{minipage}
\end{table}

 The next step consisted on a cut on a neural network (NN) output. 
For each neutralino mass, a NN was optimized for the simulated signal and
a background sample consisting of simulated atmospheric muons.
 No information about the characteristics of atmospheric neutrinos was
 included in the training of the network. 
Eight variables were used as an input to the NN to characterize the events: the number of hit
channels, the difference in reconstructed angle between the fast first-guess fit and 
the maximum likelihood fit, the number of 'late' ($t_{\it res}>$150 ns) 
hits, the track length of the first-guess fit, the length spanned by
all the hits projected onto the direction of the final reconstruction, the likelihood
of the fit, the center of gravity of the hits and the vertical distance
between the deepest and shallowest hit modules. The MLPfit package was used~\cite{MLPfit}. 
The network performance was optimized by varying the network architecture, the
 size of the training samples and minimization algorithms.
 An architecture of 8-40-1 with a hybrid linear minimization was found
 to give maximal efficiency. \par
\begin{figure}[t]
\centering\epsfig{file=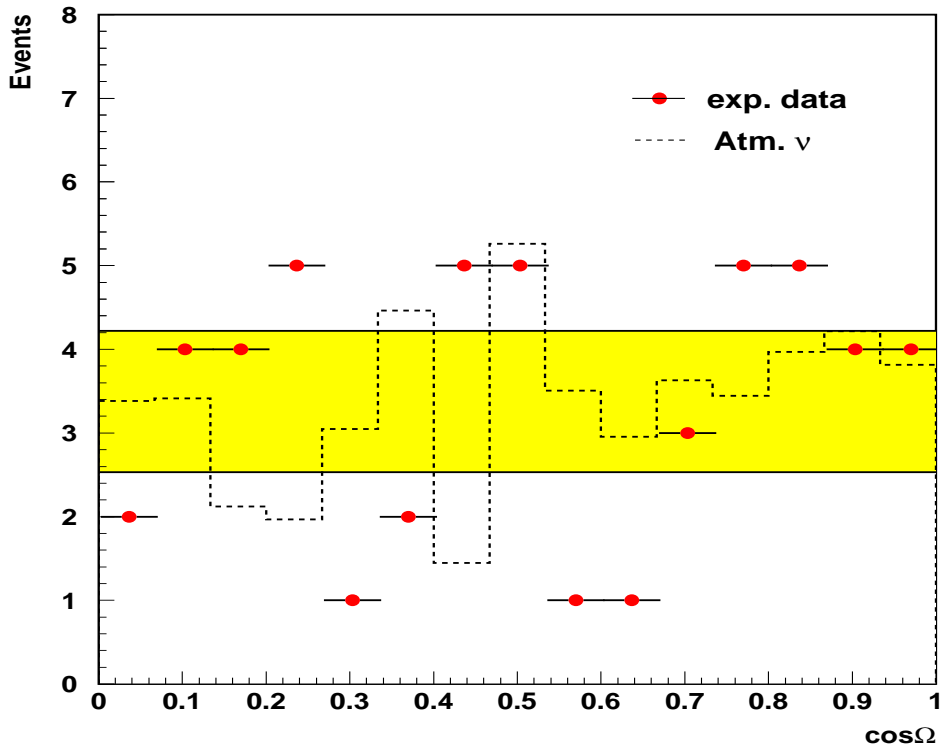,height=0.8\linewidth,width=\linewidth}
\caption{Cosine of the space angle between the reconstructed track
  and the Sun position at final cut level for data and atmospheric neutrinos. The case shown is for
  the cut optimization for 500 GeV neutralinos (annihilation into $W^+W^-$). 
 The shaded area shows the one standard deviation systematic 
  uncertainty in the expected atmospheric neutrino distribution.}
\label{fig:delta_Phi}
\end{figure}

 The final event selection consisted of the following series of cuts: the number of hit channels, the
 number of 'early' hits ($t_{\mbox{\tiny res}}<$ -15 ns), the number of
 hits used in the first-guess fit and the distance between the first and last direct hits projected on
 the track direction. For this last cut, the direct hit definition used was  
 -15\,ns $< t_{\mbox{\tiny res}} <$75\,ns. 
Figure~\ref{fig:relative_effs_L5} shows an example of the
 relative efficiency for signal, background and data as a function of cut
 level, for one of the neutralino masses. Cut level 4 corresponds to
 the  cut on the NN output, and the last level to the addition of the cuts just described, shown 
individually as the intermediate points between level 4 and 5 in the plot. 
Figure~\ref{fig:delta_Phi} shows the distribution of the cosine of the space angle, $\Omega$, between the
reconstructed tracks and the direction of the Sun (after unblinding). The  distribution is in
agreement, both in shape and 
 normalization, with the expectation from the atmospheric neutrino background, within the 
25\% overall systematics of the atmospheric neutrino flux~\cite{Gaisser:02a} (indicated by the
shaded band).

\section{\label{sec:uncertainties}Systematic uncertainties}
 There are several systematic uncertainties in the detector effective
 volume that should be taken into account when calculating limits. 
The uncertainty in the optical module sensitivity contributes about 14\%, 
while the detector calibration and the hardware simulation contribute
 about 5\%. The largest contribution comes from the imprecise  
 knowledge of the layered structure and the impurity contents of 
 the ice, and is less than about 30\% (with the exception of the softest 
 spectrum, annihilation of 100 GeV 
neutralinos into $b\bar{b}$ where this uncertainty reaches about 50\%). 
Uncertainties on the neutrino-nucleon cross-sections and in the simulation of muon energy loss  
  contribute of the order of 5\%. The overall systematic
 uncertainties in the detector effective volume lie between 20\% and about 60\% depending on the neutralino
 mass model being tested (see~\cite{Yulia} for details). The uncertainties have been
 incorporated in the limits using the 
 method described in~\cite{Conrad} with the unified
 Feldman-Cousins ordering scheme~\cite{Stuart:91a}.\par

\section{\label{sec:results}Results}
 
Different angular bins around the position of the Sun for each neutralino mass and
 annihilation channel were used in the search for a signal. The bins were chosen as containing 90\% of the
 signal, and they range from 5.5$^{\circ}$ for the
 harder neutrino spectra to 26$^{\circ}$ for the softest spectrum 
(100 GeV neutralinos annihilating into $b\bar{b}$). The off-source data
 in the declination band of the Sun were used to estimate 
the expected background at the Sun position. This procedure eliminates any effects of  
uncertainties in the background simulation and in the total normalization of the atmospheric 
neutrino flux. 
Table~\ref{tab:events_at_L5} summarizes the relevant numbers. \par

No evidence of a statistically significant excess of events from the direction of the Sun was found. 
Therefore, the number of observed events from that direction, together with the 
Poisson mean of the background expectation, were used to set limits on the neutrino flux at the Earth from 
neutralino annihilations in the Sun. We have followed the
 same prescription as in~\cite{AMANDA:02a}, starting from the directly
 measurable quantity, the neutrino-to-muon conversion rate,
 $\Gamma_{\nu\rightarrow\mu}\,=\,\mu_{90}/(V_{\mbox{\tiny eff}}\,\epsilon\,t)$,
where $\mu_{90}$ is the 90\% CL signal upper limit, $t$ is the live-time, $\epsilon$ the
 reconstruction efficiency and $V_{\mbox{\tiny eff}}$ the effective volume of the detector at final cut level. 
The limit on the annihilation rate of neutralinos in
 the Sun, $\Gamma_{\mbox{\tiny A}}$, is proportional to
 $\Gamma_{\nu\rightarrow\mu}$, where the proportionality coefficient takes into account  
the production of muons through the neutrino-nucleon cross-section weighted by the different branching ratios 
of the $\widetilde{\chi}^{\tiny 0}_{\tiny 1}\widetilde{\chi}^{\tiny 0}_{\tiny 1}$ annihilation process and the 
corresponding neutrino energy spectra. From $\Gamma_{\mbox{\tiny A}}$ we reach the
 limit on the muon flux at the Earth by taking into account that 
\begin{equation}
\phi_\mu(E_\mu \ge E_{\mbox{\tiny thr}})\,=\,
\frac{\Gamma_{\mbox{\tiny A}}}{4\pi D_{\odot}^2}
\int_{E_{\mbox{\tiny thr}}}^{\infty} dE_\mu
\frac{dN_\mu}{dE_\mu},
\label{eq:mu_flux1}
\end{equation}
where D$_\odot$ is the distance to the Sun and $dN_\mu / dE_\mu$ is
 the muon flux produced from the neutralino annihilations, 
 including all MSSM model dependencies and interaction kinematics. 
$E_{\mbox{\tiny thr}}$ is here an arbitrary threshold, that can
 be used to convert the measured flux limit (which is obtained for the
 actual detector energy threshold) to any other threshold. 
 The last three columns of table~\ref{tab:events_at_L5} show the Feldman-Cousins
  90\% upper limit on the expected signal and the corresponding 90\% CL limits on the 
annihilation rate at the center of the Sun and on the muon flux at the
 Earth. All~numbers~include~systematic~uncertainties.\par

\begin{figure}[t]
\centering\epsfig{file=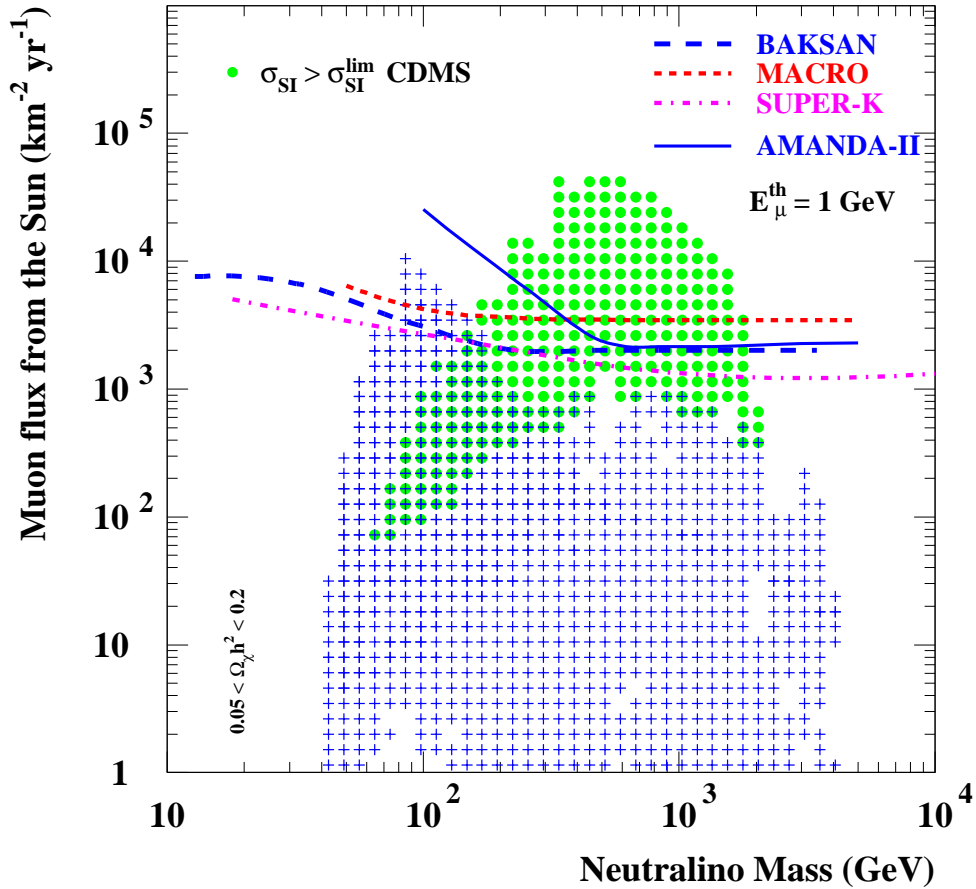,height=\linewidth,width=\linewidth}
\caption{Upper limits on the muon flux from the Sun, $\Phi_{\mu}$, from neutralino 
  annihilations into $W^+W^-$ (hard channel) as a function of neutralino mass. The muon
  energy threshold has been extrapolated to a common value of 1 GeV. 
The dots show the models disfavored by  recent direct search results from CDMS.}
\label{fig:limits}
\end{figure}

\section{\label{sec:outlook}Discussion}

 Figure~\ref{fig:limits} shows the AMANDA limits on the muon flux at the Earth 
 from neutralino annihilations into $W^+W^-$ (hard channel) in the Sun as a function of neutralino
 mass, compared to the results of  Baksan~\cite{Baksan},
 MACRO~\cite{MACRO} and Super-K~\cite{SuperK02} 
 and theoretical predictions based on {\texttt{DARKSUSY}}.
The limits have been rescaled to 
a common muon threshold of 1~GeV using the known energy spectrum of the neutralinos. 
The sparseness of the arrangement of the AMANDA
 strings (not less than 25 m between nearest neighbours) makes the detector less effective at lower
 neutrino energies, and leads to a worsening of the limit for
 neutralino masses below about 500 GeV. Above this mass, AMANDA, with the 
144 days of live-time used for the present analysis, is already 
 competitive with experiments with much longer exposure times (3.6 years of
 collected data for MACRO and 4.6 years for Super-K for example). \par
 The symbols in the figure represent the predictions of {\texttt{DARKSUSY}}  for
 given combinations of  SUSY parameters. The models shown are only those
 which give a relic density within the current limits of allowed 
dark matter density in the universe, 0.05$< \Omega h^2 <$ 0.2. 
The dots represent parameter combinations that
 are disfavored by the latest results from direct experiments setting
 limits on the neutralino-nucleon cross-section~\cite{CDMS}. 
Although comparison between direct and indirect
 searches is not straightforward, the figure shows that current results from indirect
 searches are competitive in the high neutralino mass region.\par
 Neutrino oscillations in the atmosphere (not included in the simulations) would tend to reduce 
the muon-neutrino flux and affect the total normalization of the background. Since we have used 
the measured neutrino flux off-source as an estimation of the background at the source
position, the effect of oscillations is inherently included in the analysis. 
The atmospheric neutrino simulations were used as a
confirmation that the analysis cuts selected atmospheric neutrinos and rejected the down-going
atmospheric muons to the desired level before the unblinding of the data set. \par
 The {\texttt{DARKSUSY}} simulations did not include oscillations 
of the neutrinos on their way to the Earth. 
The effect of oscillations of neutrinos produced in the Sun on the final 
flux observed at the Earth has been discussed in~\cite{Marek:01a} (and in~\cite{Crotty:02a} in a 
slightly different context). 
Detailed calculations in~\cite{Marek:01a} found that the resulting muon flux at the Earth from low
mass ($m_{\widetilde{\chi}^{\tiny 0}_{\tiny 1}} <m_t$) MSSM neutralino annihilations can be considerably affected 
by $\nu_\tau \leftrightarrow \nu_\mu$ oscillations in some scenarios. This is the case for MSSM models 
with important annihilation branching ratios into  $\tau^+\tau^-$, typically $B_{\tau\tau}> 0.1$. In such cases the 
muon flux at the Earth can be increased by a factor of about 2 to 4 with respect to the 
no-oscillation case. 
For higher neutralino masses, where the main annihilation channel can be through gauge bosons or heavy quarks, 
the effect of  $\nu_\tau \leftrightarrow \nu_\mu$ oscillations is negligible, 
 oscillations into $\nu_\tau$ being compensated by $\nu_\tau$ oscillating into  $\nu_\mu$.
Flavor mixing with $\nu_e$ both from  $\nu_\mu$ and $\nu_\tau$ is suppressed by the small value of $\Delta m^2_{21}$ 
and the mixing angle $\sin^2 2\theta_{31}$ respectively, and does not play a role when considering 
the $\nu_\mu$ flux from the Sun. Given the considerations above, the flux and annihilation rate
limits presented in this letter for the case of 
annihilations into $W^+W^-$ would be practically unaffected if
one were to consider oscillations. The results for the $b\bar{b}$ annihilation channel 
would also be unaffected for high neutralino masses ($m_{\widetilde{\chi}^{\tiny 0}_{\tiny
 1}}\gtrsim 200$ GeV) and would  worsen by a few percent for  neutralino masses below $200$ GeV.\\

\noindent {\it Acknowledgements:} 
{\small We acknowledge the support from the following agencies:
 The U.S. National Science Foundation, 
the University of Wisconsin Alumni Research Foundation, 
the U.S. DoE, 
the U.S. NERS Computing Center, 
the UCI AENEAS Supercomputer Facility, 
the Swedish Research Council, 
the Swedish Polar Research Secretariat, 
the Knut and Alice Wallenberg Foundation (Sweden), 
the German Federal Ministry of Education and Research, 
the Deutsche Forschungsgemeinschaft,
the IWT (Belgium), 
the FWO(Belgium),
the FNRS (Belgium)
and the OSTC (Belgium).
 D.~F.~Cowen acknowledges the support of the NSF CAREER
 program. M. Ribordy acknowledges the support of the SNF (Switzerland)
}

\end{document}